# Is the proposed SI revision, according to the October 2018 BIPM documents, scientifically and formally satisfactory?

F. Pavese
Torino, Italy

**Abstract**

The use of the fundamental constants for the definition of the most important measurement units of the International System, was considered a good solution to found it on more solid bases. From further analysis, this solution was found implying a number of consequences under-evaluated by the BIPM until the proposal presented for discussion in November 2018 to the CGPM, or at least never clearly explained to the signatories Countries of the Metre Treaty. This lack of clarity will affect the implementations of the revised System in the future.

Following previous illustrations, this paper is focussing on some issues of direct impact on the correctness of the new definition and on its implementation: how many digits, supported by the latest experimental values, can be stipulated for the numerical values of the constants; why the present experimental uncertainties do not support the pretended precision of the constants; how inconsistencies affect some constant final database; why the CODATA LSA analysis alone is insufficient to support the stipulated constants' precision; how the need should be explained of keeping the former base units, in order to preserve their present magnitudes; which is the true base-units/constants relationship; how the now-installed hierarchy between the constants and the base units makes changes in the future metrological pyramid; how to still use the present top national standards in future.

## 1. Introduction

Pending the date when the CGPM will examine the BIPM proposal for the revision of the SI, the analysis that follows is based on the 2018 status of the structure of the revised SI at the eve of the CGPM 26[th] Meeting, on the results of the CODATA 2017 adjustment as reported in [1] and as published on some papers of the 2018 issues of Metrologia [2–4, 18], particularly on what is related to Planck constant (mass unit), $h$. In [2] a deeper analys is provided than in [1] in support of the results of the 2017 CODATA adjustment, with further details on the method used. In particular, in its Figure 2 the 2014-2017 data for the Planck constant are shown, as obtained by the CODATA. See [4–12] for previous analyses on the revision of the SI.

Paper [3] is musing on the data for the Planck constant—one of the main reasons for the commonly-agreed urgency of the SI revision—and in [4] a different analysis is performed of the same Planck constant numerical value, with diverging conclusions and it was later supplemented in [18] with a suggestion for solving the present inconsistency affecting the mass unit.

Table 1a,b reports here for convenience the CODATA values of the constants since 2006.

Table 1a. Change in numerical value and its uncertainty *u* of the CODATA adjustments 2006-2017, for the new constant *k* involved in the revised SI definition. For *e*, *k*, $N_A$ in Table 1b, see the Appendix.

| Constant | CODATA | Numerical value * | $u$ / relative $\times 10^7$ | Change | Total shift |
|---|---|---|---|---|---|
| *Planck* $h \times 10^{34}$ | 2006 | 6.626 069$_0$ | 3.3/0.50 | — | |
| | 2010 | 6.626 069$_6$ | 2.9/0.44 | $6 \times 10^{-7}$ | |
| | 2014 | 6.626 070 04$_4$ [a] | 0.8/0.12 | $4.4 \times 10^{-7}$ | $10.4 \times 10^{-7}$ |
| | 2017 | 6.626 070 1$_5$ [a] | 0.7/0.11 | $1.1 \times 10^{-7}$ | $11.5 \times 10^{-7}$ |

* The CODATA reports standard uncertainty *u*. The smaller-case digits, obtained from the CODATA two-digit uncertainty format, are those affected by *uncertainty*.
[a] The 2014 CODATA outcome is 6.626 070 040(81), therefore the numerical value can be as low as 6.626 069 959, thus affecting also the preceding digit. Similarly for the 2017 one, the CODATA outcome is 6.626 070 150(69). Consequently the upper bound of the 2017 interval is 6.626 070 121, and the lower bound of the 2017 interval in 6.626 070 081, *not significantly overlapping*.

Table 1b. Change in numerical value and its uncertainty *u* of the CODATA adjustments 2006-2017, for the new constants *e*, $N_A$ and *k* involved in the revised SI definition. See Table 1a for *h* in the main text.

| Constant | CODATA | Numerical value * | $u$ ($k = 1$)/ relative $\times 10^7$ | Change | Total shift |
|---|---|---|---|---|---|
| *Electron charg* $e \times 10^{19}$ | 2006 | 1.602 176 4$_9$ | 0.4/0.23 | — | |
| | 2010 | 1.602 176 5$_7$ | 0.35/0.19 | $8 \times 10^{-8}$ | |
| | 2014 | 1.602 176 6$_2$ | 0.1/0.06 | $5 \times 10^{-8}$ | $1.3 \times 10^{-7}$ |
| | 2017 | 1.602 176 634 [b] | 0.08/0.05 | $1.4 \times 10^{-8}$ | $1.4_4 \times 10^{-7}$ |
| *Avogadro* $N_A \times 10^{-23}$ | 2006 | 6.022 141$_8$ | 3.0/0.50 | — | |
| | 2010 | 6.022 141$_3$ | 2.7/0.45 | $-5 \times 10^{-7}$ | |
| | 2014 | 6.022 140$_8$ [c] | 0.7/0.12 | $-4 \times 10^{-7}$ | $-9 \times 10^{-7}$ |
| | 2017 | 6.022 140 7$_6$ [c] | 0.6/0.10 | $+1 \times 10^{-7}$ | $-8 \times 10^{-7}$ |
| *Boltzmann* $k \times 10^{23}$ | 2006 | 1.380 650$_4$ [d] | 24/17 | — | |
| | 2010 | 1.380 648$_8$ [d] | 13/9.4 | $-1.6 \times 10^{-6}$ | |
| | 2014 | 1.380 648$_5$ | 8/5.8 | $-0.3 \times 10^{-6}$ | $-1.9 \times 10^{-6}$ |
| | 2017 | 1.380 649 [e] | 5/3.6 | $+0.5 \times 10^{-6}$ | $-1.4 \times 10^{-6}$ |

* The CODATA reports standard uncertainty *u*. The smaller-case digits are *uncertain*, taken from the CODATA two-digit uncertainty format—except for Boltzmann constant *k*, see note c) and e).
[b] The 2017 CODATA outcome is 1.602 176 6341(83), therefore the numerical value can be as low as 1.602 176 6258 and as high as 1.602 176 6424, thus affecting also the preceding digit.
[c] The 2014 CODATA outcome [13] is 6.022 140 857(74), therefore the numerical value can be between 6.022 140 783 and 6.022 140 931, thus affecting the preceding digit. The analysis is similar for the 2017 one, where the CODATA outcome is 6.022 140 758(62).
[d] Two digits are shown because the rounding affects also the preceding digit.
[e] The CODATA 2017 outcome is 1.380 649 03(51), thus the rounding does not include uncertain digits (the only occurrence in the Table). However, the numerical value can be as low as 1.380 648 50, thus in fact affecting the last digit.

## Analyses and Comments

### *"Exactness" of the CODATA stipulated data*

Concerning papers [1, 2], a reader informed on the pending SI-revision process notes that the number of digits now proposed by the CODATA for the stipulated constants, possibly with the exception of the Boltzmann one, $k$, is larger than previously aimed at. That is certainly due to the lowering of the experimental uncertainties—in the period since 2006 an outstanding ≈5 times—but, apparently, it is also due to a particular use of the original data and of their associated uncertainties. Table 2 a,b shows both the CODATA proposed/stipulated values and "exactly-known values", when affected both, by an uncertainty as in Table 1 (a) and instead (b) by an *expanded uncertainty*—common practice in science for very important issues—because its use here should be considered mandatory.

Table 2 a. Different ways to treat the digits of the same numerical values of the four constants ($k = 1$).

| Constant numerical value | CODATA 2017 ($k = 1$) [1] | CODATA–stipulated [b] [1] | Exactly-known number (*rounded*) | Exactly-known number (*truncated*) |
|---|---|---|---|---|
| $\{h\} \times 10^{34}$ | 6.626 070 150(69) | 6.626 070 15 | 6.626 070 | 6.626 070 |
| $\{e\} \times 10^{19}$ | 1.602 176 6341(93) | 1.602 176 6$_{34}$ | 1.602 1766 | 1.602 1766 |
| $\{N_A\} \times 10^{-23}$ | 6.022 140 758(62) | 6.022 140 7$_{6}$ | 6.022 141 | 6.022 140 |
| $\{k\} \times 10^{23}$ | 1.380 649 03(51) | 1.380 6$_{49}$ | 1.380 65 | 1.380 64 |

Table 2 b. Different ways to treat the digits of the same numerical values of the four constants ($k \approx 2$).

| Constant numerical value | CODATA 2017 ($k \approx 2$) [a] | CODATA–stipulated [b] [1] | Exactly-known number [c] (*rounded*) | Exactly-known number [c] (*truncated*) |
|---|---|---|---|---|
| $\{h\} \times 10^{34}$ | 6.626 070 150(138) | 6.626 070 15 | 6.626 070 | 6.626 070 |
| $\{e\} \times 10^{19}$ | 1.602 176 6341(186) | 1.602 176 6$_{34}$ | 1.602 1767 | 1.602 1766 |
| $\{N_A\} \times 10^{-23}$ | 6.022 140 758(124) | 6.022 140 7$_{6}$ | 6.022 141 | 6.022 140 |
| $\{k\} \times 10^{23}$ | 1.380 649 03(102) | 1.380 6$_{49}$ | 1.3807 | 1.3806 |

[a] *Three* digits are left here for uncertainty only to allow appreciating the difference with respect to the normal CODATA 2017 two-digit estimate in the previous Table.
[b] The smaller-case digits are used here for those affected by uncertainty in the previous column.
[c] Here "exact" means unaffected by the original experimental uncertainty interval.

The CIPM preference expressed in [15], also shared by CCU, was "*for the minimum number of digits [of the stipulated value] for each defining constant h, e, k, and $N_A$ of the revised SI that yields consistency factors equal to* 1 *within their uncertainties*". The CODATA-proposed stipulations [1, 2] in Table 2 are intended to match it.

However, there is *another principle*, explicit in the CIPM/CCU rules when they talk of "consistency factors", which must also be respected in stipulation: the "continuity principle" in the magnitude of the base units. Values of the "consistency factors" (equaling 1 exactly for perfect consistency) are computed here in Eqs. (1):

$$[M(K)/(kg)_{rev}]/1 = 1.000\,000\,001(10) \qquad [1.2\times10^{-8}]$$
$$[\mu_0/(H\,m^{-1})_{rev}]/(4\pi\times10^{-7}) = 1.000\,000\,000\,20(23) \qquad [2.3\times10^{-10}] \qquad (1)$$
$$[M(^{12}C)/(kg\,mol^{-1})_{rev}]/0.012 = 1.000\,000\,000\,37(45) \qquad [4.5\times10^{-10}]$$
$$[T_{TPW}/(K)_{rev}]/273.16 = 1.000\,000\,01(37) \qquad [5.7\times10^{-7}]$$

These numerical values are based on the CODATA *stipulated* values obtained from the *adjusted* values in Table 1, and were considered to just correspond to the CIPM indicated criterion. In the Resolution A proposed by BIPM to CGPM [17], the above four relative uncertainties become: $1.0\times10^{-8}$ for $M(K)$, $2.3\times10^{-10}$ for $\mu_0$, $4.5\times10^{-10}$ for $M(^{12}C)$ and $3.7\times10^{-7}$ for $T_{TPW}$: however this indication is *false*, not being supported by the experimental data, as shown above in Table 2, by at least one order of magnitude—$10^{-7}$ instead of $10^{-8}$ and $10^{-9}$ instead of $10^{-10}$.

In fact, there is a basic difference between the *scientific* context (the constants, CODATA) and a *regulatory* context (the SI, CIPM, Metre Treaty).

Eqs. (1) may satisfy the CIPM requirement [15] that the *continuity principle* is satisfied within the present-SI best realisation uncertainties, but they are *not* supported by the *actual* experimental evidence, as shown in Table 1—and in Table 2.

A **basic dilemma** thus arises from the above situation, so far apparently undetected, certainly unpublished and unresolved, for the constant numerical values—based on the measurements performed with the *present-SI*:

**(0)** *Accepting constant stipulated values exceeding the present experimental accuracy, but preserving unit magnitude continuity,* [the above BIPM solution] or,

**(1)** *Requiring constant stipulated values to conform present experimental accuracy, but inducing a unit magnitude discontinuity* (small but significant). [see the above Tables]

That **unresolved conflict** (see also last section) consists of the fact, rather obvious, that one should not stipulate a number with more digit(s) than those *confirmed exact* by the experiments. The fact that the uncertainty will eventually be dropped in stipulation is totally irrelevant: uncertainty means that, digit(s) affected by it could be *presently* different from the stipulated one(s).

It is a fact that the present experimental results still do not support a firm continuity of the units' magnitudes. To get the desired continuity one is obliged to violate the existing support, and 'guess' the value of the last digit(s) affected by uncertainty.

*Inconsistent data and their effect*

Another major issue was raised since the 2017 CCU document [15], concerning the evident inconsistency of several supplied new data for the Planck constant: "*… Notes … that work is under way in NMIs to understand the cause for the dispersion of the experimental determinations of the Planck and Avogadro constants …*"—as also noted by the CODATA in [1-2]. Nevertheless, the CCU concluded "*that numerical values and uncertainties for the Boltzmann constant and the Avogadro constant provided by the CODATA Task Group on Fundamental Constants in their special Least-Squares Adjustment of the experimental data provide a sufficient foundation to support the redefinition, …*" and recommended the CIPM to proceed for the 26th CGPM in 2018. The CIPM did so in its 2017 meeting [14].

However, the CCM, in its 2013 Resolution, [18] had required, for accepting a stipulated value of $h$ "*at least three independent experiments, including work from Watt balance and XRCD experiments, yield consistent values of the Planck constant with relative standard uncertainties not larger than* $5 \cdot 10^{-8}$", and that "*at least one of these results should have a relative standard uncertainty not larger than* $2 \cdot 10^{-8}$".

That condition is *not so far fulfilled*, and that was recognized by the CCM in its 2017 recommendation: "*that most recent measurement results with relative standard uncertainty below* $5 \cdot 10^{-8}$ *do not pass the standard chi-squared test of consistency, but it is expected that the CODATA value and uncertainty for the Planck constant will be suitable for even the most demanding applications*" [19] and proposed "*that the results of an ongoing key comparison are used for a correction of future measurement results, to ensure the consistent dissemination of the redefined kilogram*" (emphasis added). [18]

If data inconsistency is called, as by default, an evidence of non-overlapping uncertainty intervals (for $k = 1$ in case of the CODATA) for the data, Fig. 2 in [2] shows such a case for three 2017 data, which have been considered as such also by the CODATA. In the latter respect, the conclusion in [3], based on different specific statistical tools, does not support the lack of inconsistency. On the other hand, in [4] evidence for inconsistency comes again from the use of a different analysis method of the 2017-available data. Those three 2017 inconsistent data are directly used in the analysis in [4] bring to important and conspicuous consequences: while the CODATA 2017 adjusted value for $h$ results basically equal to NRC-17, from Figs. 3-4 in [4] (to be compared with Fig. 2 in [3]) the evidence comesthat the continuing trend towards higher values of $h$—see also the above Table 1—points rather to the IAC-17 value—not consistent with the CODATA 2017 adjusted value [2].

The drift trend in 2017 is still sufficiently significant to allow the doubt that the CODATA conclusion—and the assertion in [3]—are not sufficiently founded. In fact, Figs. 3-4 in [4] also show an increase of the credible interval, another reason for being cautious about the number of stipulated digits. These facts brought to the conclusion in [4] that: "*Although nothing can be concluded about a possible future*

*development of the CODATA values for the Planck constant, their contingent change over the past decades does not encourage a redefinition of the kilogram at present*".

*CODATA treatment of inconsistent data*

In [1] it is said: "… *To achieve consistency, multiplicative expansion factors were applied to the uncertainties … The uncertainties of these input data are multiplied by a factor of* 1.7. *With this expansion of the uncertainties of the eight data, five have relative standard uncertainties $u_r$ at or below* $50 \times 10^{-9}$, *with two at or below* $20 \times 10^{-9}$ …"—facts also indicated in [2], where it is specified: "*It is note worthy that even after applying an expansion factor of* 1.7 *to the uncertainties of all … data, thereby bringing them into agreement, the relative uncertainties of the first five values of h … are, in parts in* $10^9$, *only* 15, 20, 23, 34, *and* 42, *respectively*".

The reported uncertainty lowering is strictly a feature of the *consistency-checking* LSA method. Here, it also shows its weakness for application in cases like this one.

Actually, it is certainly not the first time that the uncertainty of a constant is reduced thanks to the understood connections that the LSA method is establishing between all the elements of the dataset. In this case, it might indicate that the effect of the dispersion of the 2017 values for *h*, after having been assigned a 1.7-larger uncertainty, becomes almost irrelevant for the general consistency-degree of the whole dataset (Note that here consistency has a different meaning with respect to the data consistency, as discussed in [3, 4] by using specific statistical tools different from the CODATA one). However, since here the uncertainty lowering does not reflect onto the experimental findings, it should be considered as a LSA artefact, and the conclusions reported in [4] valid.

Further, the CODATA method of increasing the data uncertainty to eliminate the inconsistencies is common in metrology: however, it should be considered as a better-than-nothing solution, since the discrepancy could be, in reality, not due to an under-estimate of the uncertainty, but to a bias of the provided value. In the present case, it is not without inconveniences. The fact that the uncertainty after its 'expansion' remains nearly the same of that in 2014 is not necessarily good news, but it indicates instead that the 2017 values were in practice brought to have an irrelevant effect, so that the 2017 adjustment coincides in practice with the 2014 one. The numerical value could have become impaired by the small sensitivity of the 2017 constant's subset with respect to the overall dataset. As a consequence, it may have happened that the value was adjusted more or less than correctly.

**Insufficient overall analysis of the used database**

In conclusion, the confidence/degree-of-believe on future stability of the numerical values of *h* should be considered insufficient. A deeper discussion of the evident inconsistencies of several data should be provided.

The analyses in [2–4], contrasting with each other, are presently insufficient to draw the conclusion of sufficiency about such an important subject matter. In particular, the CODATA 2017 value of *h* is not supported by all presently-published analyses. Considering the extraordinary effort made by several NMIs for supplying more data in order to support the decisions that must be taken about the numerical values to assign to the constants, one would expect that deeper analyses are made available in support to the results of the 2017 CODATA adjustment, and, beyond it, to the available dataset—which should possibly be increased by means of a *stand-by time of a couple of years before the revision comes into effect*.

The importance of the result of the SI revision, not only for the metrologists but for the entire Community of scientists, should prompt a broad number of competent and independent analyses, using different methods. In this respect, also analyses independent on the CODATA one should be included.

In general, the LSA method allows checking only *consistency* of the dataset, because the measured values are changed ("adjusted") to *optimise the standard deviation* of the set. This method, sound for many scientific applications, looks unsuitable when, as for the SI, the *numerical values* of the constants are instead the unique goal: the values supplied by CODATA are *relative to the constraints* chosen to make determined LSA relational-equation set— strictly speaking, the LSA is not a statistical method for obtaining "best" mean values of the dataset.

The advantage to use more methods would be to mitigate the otherwise un-confronted effect of the somewhat-biased values and reduction of uncertainty levels caused by the CODATA use of the LSA, so leading to better evidence about the digits needed and allowed to express the numerical values of the constants—and the Planck one in particular. In turn, that would offer higher confidence to the process of stipulation of "exact" numerical values. A combined "best value", and its associated uncertainty, should be obtained by using several diverse methods.

In [18] a suggestion is made for implementing the CCM 2017 resolution [19], to treat the available dataset of future KC comparisons with a fixed-effect statistics—a random effect statistics was not found sufficient for the purpose. That method is basically similar to the one used by CODATA for their value adjustment. However, that type of method only allows to firmly find *the pair-differences* between the data, each value remaining instead undetermined by a fixed constant, generally unknown. Therefore, that method is not suitable for providing the required "*corrections*" of the comparison results, unless the stipulated value of the constant in question is assigned to a "reference laboratory", *which cannot be anybody else except the BIPM itself, due to its international status*.

**Is a hierarchy between countries now established, or**
**are the present top national standards still valid after the SI revision?**

This issue, non-scientific in itself but basic in the SI regulatory context, is fully discussed in [12]. Here only some conclusions are reported.

The metrological **traceability** pyramid of the standards is changed in the revised SI, as shown in Table 3: "definitional methods" do not stand anymore, but "primary methods", *not* to be included among those of the *mise en pratique*, as CIPM is still doing, should be identified and replace them. For the standards below the latter level nothing changes.

Table 3. Metrological traceability chain for the SI (*example: length*). (from [7])

| Traceability Level | Present-SI | | Revised-SI |
|---|---|---|---|
| Top | Definitional method *Method using "distance" and interval"* [a] | *"time"* | No definitional method. *"Condition"*: to reproduce the stipulated constant(s) value |
| −1 | Mise en pratique. *Other method(s): using frequency and period* | | Primary methods: $c_0$ *and t explicit in the model* |
| −2 | Secondary methods. *Other method(s): stabilised laser* | | Mise en pratique. *Other method(s): using frequency and period* |
| −3 | Workshop methods. *Gauge blocks* | | Secondary methods *Other method(s): stabilised laser* |
| −4 | Lower ranks | | Workshop methods: *gauge blocks, …* |
| … | … | | |

[a] This requirement is so far not always respected in the traceability chain, using instead frequency and period.

In addition, while at present, the implementation of the SI according to the Metre Treaty, in particular by the NMIs, never implies that they must resort to another NMI/Country, so that a user might decide to resort to another NMI/Country *only on its own choice*, with the revised SI definition, traceability to the definition requires the demonstration that the defined values of the constants are *determined* by the NMIs. This is *affordable only by a few Countries*, unless reference to the constants becomes only a *check*, not **the** SI *definition*, differently from the presently proposed definition.

If this change will not be implemented, a **hierarchy** between Countries will necessarily be established between those who will experimentally determine the whole set of constants and the rest— the vast majority.

On the other hand, for the present standards at the top of the traceability chain, which were used to *determine the numerical values* of the constants used in the new SI definition, a different approach can be considered, still being a controversial one.

It is a fact, as said hereinbefore, that the numerical values of the constants are those obtained by *using the units of the present-SI*. Therefore, since new and old units are made *indistinguishable in magnitude* (with the two exceptions below), one could ask why the present standards, having provided so far the numerical values to the constants, up to top metrological level, should not be anymore permitted to be used after promulgation of the revision. In fact, should, e.g., the deadline now moved to, say, 2020, they would continue to provide new valid data, and the stipulated value would be adjusted if these further data will not be consistent with the 2017 ones.

In addition, also for at least a further "short period of time" [16]—i.e. under "repeatability conditions" [16]— after the revision comes into effect, stable standards should entitled to be used for *further valid realizations* of the constants (note, *now not* providing a different numerical value). Therefore, those same numerical values— now no more in the definitions and thus uncertain—remain *by definition consistent* with the new condition set by the use of the constants. For a "short period of time" [16], here means until evidence will become available in future, from new experiments or theoretical reasons, that the present units were actually not consistent with each other.

Immediately after the change of definition (now 20 May 2019), they still *ensure the consistency* ("metrological compatibility" [16]) of the old with the new units. This means respecting the "principle of continuity", obviously within the uncertainties associated to the results obtained with the present-SI. It is an *intrinsic property of the previous standards* that is still valid and should be preserved by a clear indication in the BIPM texts.

**Relationship between Base units and Constants**

The issue is fully discussed in [8, 10]. Here only some conclusions are reported.

The SI revision was considered by the proposers to produce a scientific revolution in Kuhn's sense, so requiring a brand new approach to accommodate the changes in the SI. However, not all these changes are correctly identified in the current BIPM documents. They would be less dramatic than estimated and could be accommodated without such revolution provided that its *conceptual structure* has enough flexibility: (i) with *constants* based on the principles and tools of and contents of fundamental physics, and thus in particular on the currently accepted system of quantities and the set of fundamental constants: (ii) with (not optional) *base units* linked to the current SI so that the principle of continuity is fulfilled.

A conceptual roadmap that satisfies both these requirements, can be obtained by construing a system of units according to an explicit *two-stage structure*—implicit but not implemented in the present document— including (i) a *fundamental* system and (ii) a *conventional* one:

(i) both a system of quantities and a set of constants corresponding to the dimensions of the base of the system are assumed; this is the *fundamental* system where the *numerical value of each constant is* 1;

(ii) a *conventional* system is then considered linked to the above fundamental system, where it is admitted that the numerical values of the constants can have values different from 1, assigned according to the best available *present* knowledge, so that in changing to the new system the units maintain their values as expected according to the principle of continuity.

**Discontinuities occur in the magnitude of some of the new measurement units**

In addition to the fact that, should the support of the experimental data be maintained in the stipulation of the numerical values of the constants, the magnitude continuity condition would *no more* be satisfied by at least one order of magnitude (see hypothesis (1) above), at present, two derived units will show a magnitude discontinuity of the order of $10^{-7}$ relative, quite significant, in the revised SI proposal:

– the **volt**, for which a cause is not presently explained in any publication, but probably arises from the imperfect "closure" of the "quantum triangle";

– the **dalton** with respect to the mole, arising from the fact that in the revised SI the Dalton is now affected by an uncertainty [17] while the mole is defined exact.